\documentclass[letterpaper,twocolumn]{jpsj3}
\usepackage{txfonts}
\usepackage{comment}
\usepackage{bm}

\title{$S$-wave Superconductivity in the Dirac Line-nodal Material CaSb$_2$}

\author{Hidemitsu Takahashi$^1$\thanks{takahashi.hidemitsu.23r@st.kyoto-u.ac.jp}, Shunsaku Kitagawa$^1$,  Kenji Ishida$^1$, Mayo Kawaguchi$^1$, Atsutoshi Ikeda$^{1, 2}$, Shingo Yonezawa$^1$, and Yoshiteru Maeno$^1$}
\inst{$^1$ Department of Physics, Graduate School of Science, Kyoto University, Kyoto 606-8502, Japan \\
$^2$ Department of Physics, University of Maryland, College Park, MD 20742-4111, USA}

\abst{We performed $^{121/123}$Sb-nuclear quadrupole resonance (NQR) measurements on the superconducting (SC) line-nodal material CaSb$_2$ in order to investigate electronic properties in the normal and SC states from a microscopic point of view. In the normal state, the nuclear spin-lattice relaxation rate $1/T_1$ for the Sb(1) site, which is responsible for the line-nodal parts, is approximately proportional to temperature, indicating the conventional Fermi liquid state. From comparison with band structure calculations, it is considered that the NQR properties related to the line-nodal character are hidden because the conventional behavior originating from Fermi-surface parts away from the nodes is dominant. In the SC state, a clear coherence peak just below the transition temperature and an exponential decrease at lower temperatures were observed in $1/T_1$. These results strongly suggest that conventional $s$-wave superconductivity with a full gap is realized in CaSb$_2$. }

\begin{document}
\maketitle


Topology is one of the key concepts in recent condensed matter physics. Stimulated by the discovery of topological insulators,\cite{topoins1, topoins2} a large number of studies on topological quantum phenomena have been carried out due to fundamental research interest as well as to explore their applications. The concept of topology has also been introduced to superconductivity.\cite{toposc1}$^{-}$\cite{toposc3} The Majorana zero mode is considered to be important for applications in error-free quantum computing.\cite{kitaev}

Topological semimetals are distinct types of topological materials. Dirac and Weyl semimetals are characterized by point nodes in bulk electronic bands.\cite{dandw} Weyl semimetals are realized in systems without spatial-inversion or time-reversal symmetry. Fascinating phenomena such as ultra-high mobility, surface Fermi arc,\cite{fermiarc} and chiral magnetic effect\cite{nn} are expected to occur in these materials. 

Recently, a new type of topological semimetal, namely, line-nodal semimetals, has been discovered.\cite{lines} In these materials, increase in the nodal dimension leads to notably rich phenomena. For example, long-range Coulomb interaction,\cite{longC} a large surface-polarization charge,\cite{charge} quasitopological electromagnetic responses,\cite{topoele} and drumhead surface states\cite{drumhead} are predicted. Regarding superconductivity, topological crystalline superconductivity and second-order topological superconductivity are expected in nodal-loop materials.\cite{lineSC}
Many materials are predicted to be line-nodal semimetals without spin-orbit coupling (SOC).\cite{cu3pdn, caagx} However, their nodes usually become gapped under SOC, resulting in topological insulators or point-nodal semimetals. To preserve line nodes, an additional symmetry, such as non-symmorphic symmetries is required.\cite{mirror, nonsym} Therefore, line-nodal materials with non-symmorphic symmetry provide ideal platforms for studying novel topological phenomena and unconventional superconductivity.\cite{kobayashi}

Here, we introduce CaSb$_2$, crystallizing in the monoclinic structure with a non-symmorphic space group ($P2_1/m$, No.11, $C_{2h}^2$) as shown in Fig. 1(a). From a band structure calculation,\cite{funada} it is predicted that CaSb$_2$ has Dirac line-nodes in its bulk bands protected by the combination of screw and mirror symmetries, even with SOC. A large magnetoresistance was reported,\cite{funada} which may be the signature of topological materials.\cite{tasb} In addition, some of the present authors discovered superconductivity in this compound, which has the transition temperature of $T_{\mathrm{c}}=1.7$ K.\cite{ikeda} There are several line-nodal materials which exhibit superconductivity,\cite{banbs3} but little information on superconducting (SC) symmetry is available. Therefore, electronic properties of this compound in the SC as well as normal states deserves further investigation.

Nuclear magnetic resonance (NMR) and nuclear quadrupole resonance (NQR) are microscopic measurements used to probe the electronic state at the nuclear site through the hyperfine coupling between the nuclear spin and the surrounding electrons. In the case of conventional metals or semimetals, the nuclear spin-lattice relaxation rate 1/$T_1$ is related to the quasi particle density of states (DOS) near the Fermi energy $E_{\mathrm{F}}$, and thus, its temperature $T$ dependence can reflect the nodal structure near $E_{\mathrm{F}}$. Moreover, anomalous orbital contributions to the hyperfine coupling modified by the linear band dispersion were predicted for Dirac and Weyl fermions.\cite{dirac, wyle} In particular, NQR is suitable for investigating superconductors because it can be performed without an external magnetic field. For these reasons, NQR has been playing an important role in the study of topological materials and superconductors.

In this paper, we report the results of $^{121/123}$Sb-NQR measurements that reveal properties of the normal and SC states of CaSb$_2$. This compound has two distinct Sb sites, both forming zigzag chains along the $b$-axis. We succeeded in observing the NQR signals of the Sb(1) site, whose electrons mainly form the two-dimensional topological Fermi surfaces involving the Dirac line nodes. In the normal state, 1/$T_1T$ is almost $T$-independent. This indicates that CaSb$_2$ behaves as a conventional metal above $T_{\mathrm{c}}$. In the SC state, 1/$T_1$ shows a clear coherence peak just below $T_{\mathrm{c}}$ and an exponential decrease at lower temperatures. These results strongly suggest $s$-wave superconductivity with a full gap. 

\begin{figure}
\centering
\includegraphics[width=9cm]{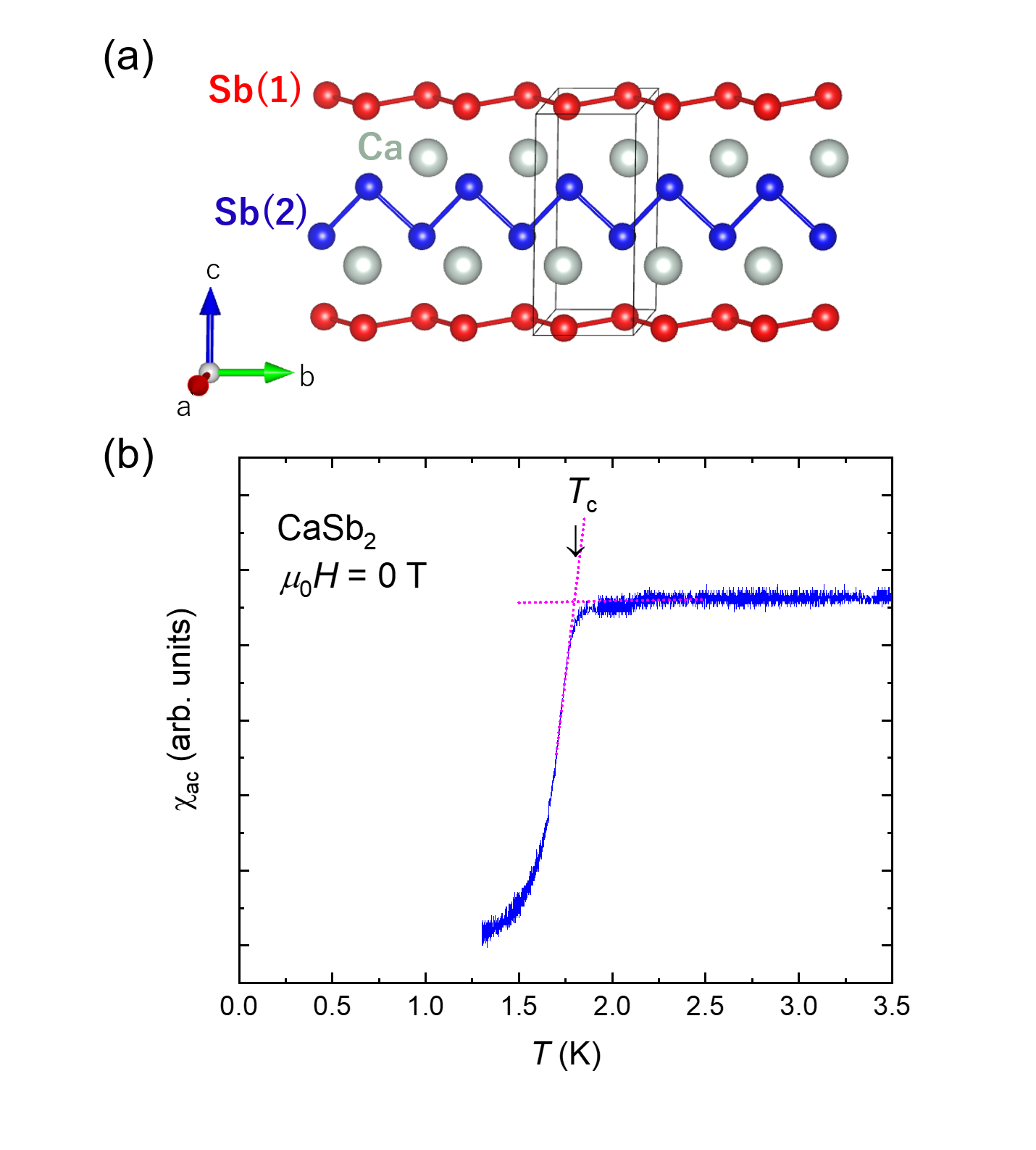}
\caption{(a) Crystal structure of CaSb$_2$ drawn by VESTA.\cite{vesta} There are two non-equivalent Sb sites both forming zigzag chains along the $b$ axis. (b) $T$ dependence of the AC susceptibility $\chi_{\mathrm{ac}}$ measured at the frequency of 75 MHz under zero field, showing a sharp superconducting transition at 1.8 K.}
\label{crystal}
\end{figure}


We used polycrystalline samples synthesized by a solid-state reaction reported previously.\cite{ikeda} The samples were powdered, packed in a plastic straw, and losely capped with epoxy. This is to make the sufficiently large surface area to enhance the intensity of NQR signals as well as to avoid excessive Joule-heating by radio frequency pulses.
Based on an AC susceptibility measurement using the NQR tank circuit, the onset $T_{\mathrm{c}}$ was evaluated to be 1.8 K, as shown in Fig.~1(b), which is consistent with the previous report.\cite{ikeda} CaSb$_2$ contains NQR-active elements $^{121}$Sb and $^{123}$Sb; Table I lists the values of their nuclear spin $I$, nuclear gyromagnetic ratio $\gamma/2\pi$, nuclear quadrupole moment $eQ$ divided by elementary charge $e$, and natural abundance (N. A.). A standard spin-echo technique was used for the NQR measurements. The value of $1/T_1$ was obtained by measuring the time dependence of the spin-echo intensity after saturation of the nuclear magnetization. A $^3$He-$^4$He dilution refrigerator was used for the measurement down to 0.3 K. 

We calculated the electronic band structure and NQR parameters of CaSb$_2$ using the full-potential linearized augmented plane wave plus local orbitals method implemented in the WIEN2k package.\cite{wien, wien2}. We adopted the Perdew-Burke-Ernzerhof generalized gradient approximation\cite{PBE} as the exchange-correlation functional. We also included the effect of SOC. The calculation was performed based on an experimental crystal structure\cite{crys} and with a $k$-mesh of $36\times39\times19$. 

\begingroup
\renewcommand{\arraystretch}{1.8}
\begin{table}
\begin{center}
\caption{Basic parameters for $^{121}$Sb and $^{123}$Sb nuclei.}
\begin{tabular}{c|cccc}\hline\hline
& $I$ & $\gamma/2\pi$ (MHz/T) & $Q$ ($10^{-28}$ m$^2$) & N. A. ($\%$)\\ \hline
$^{121}$Sb & 5/2 & 10.189 & $-0.543$ & 57.3\\ \hline
$^{123}$Sb & 7/2 & 5.5175 & $-0.692$ & 42.7\\ \hline\hline
\end{tabular}
\end{center}
\end{table}
\endgroup

\begin{figure}
\centering
\includegraphics[width=9cm]{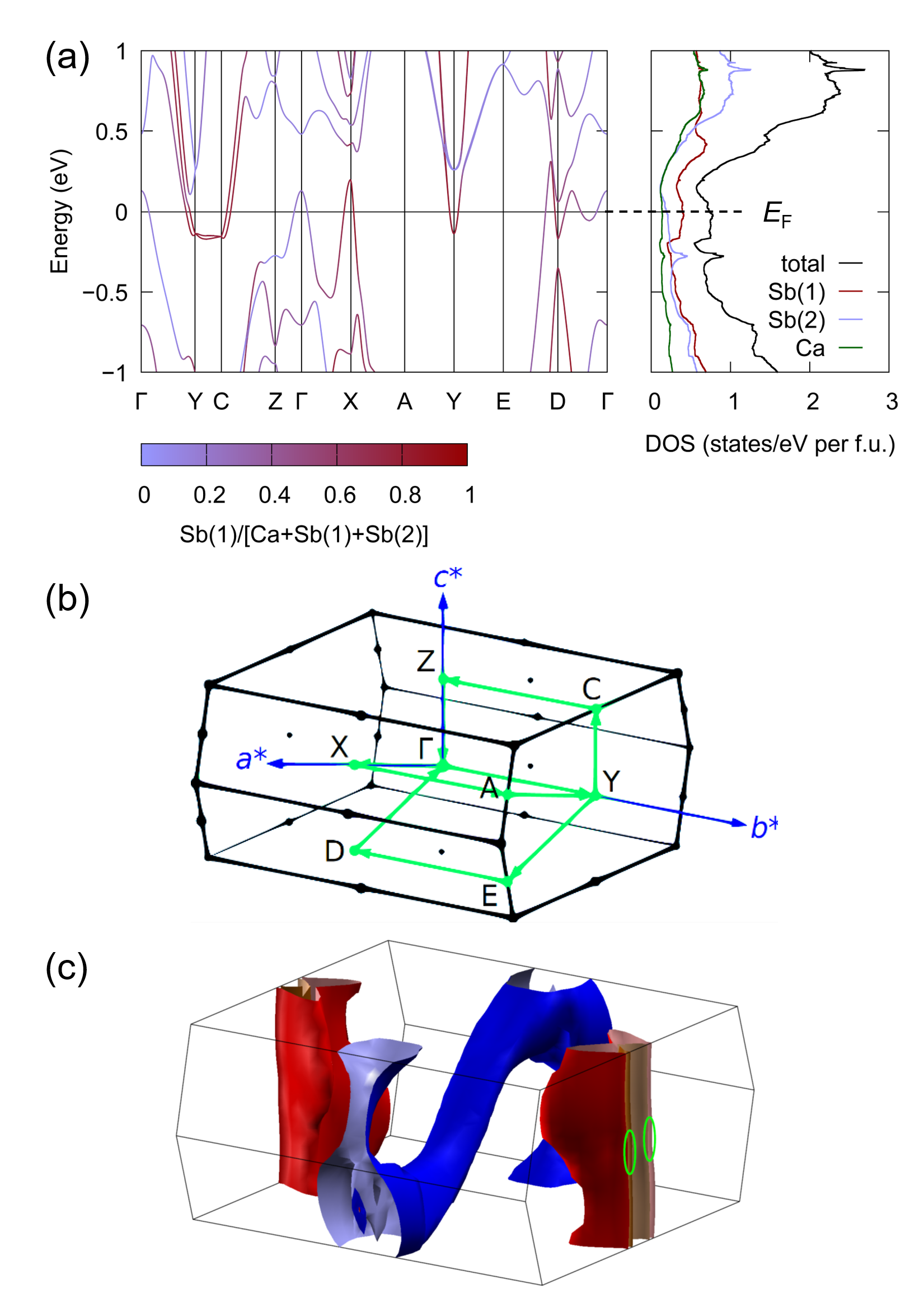}
\caption{Band structure of CaSb$_2$ calculated using WIEN2k. (a) Band dispersion showing two bands with distinct characters. The electronic DOS per formula unit (f. u.), as well as partial DOS of each site, is shown in the right. (b) Brillouin zone. The $c^*$ axis is taken as the $\mathrm{\Gamma Z}$ axis, which is different from Ref. 22. (c) Fermi surfaces. Dirac nodal lines cross the red Fermi surfaces on $b^*=\pm\pi/b$ planes. Ovals indicate the region where the nodal line cross $E_{\mathrm{F}}$ and the two Fermi-surface cylinders merge. Due to the limited resolutions of the calculation, this merging is not resolved in this figure. The blue one is a normal Fermi surface.}
\end{figure}

\begin{figure}
\centering
\includegraphics[width=9cm]{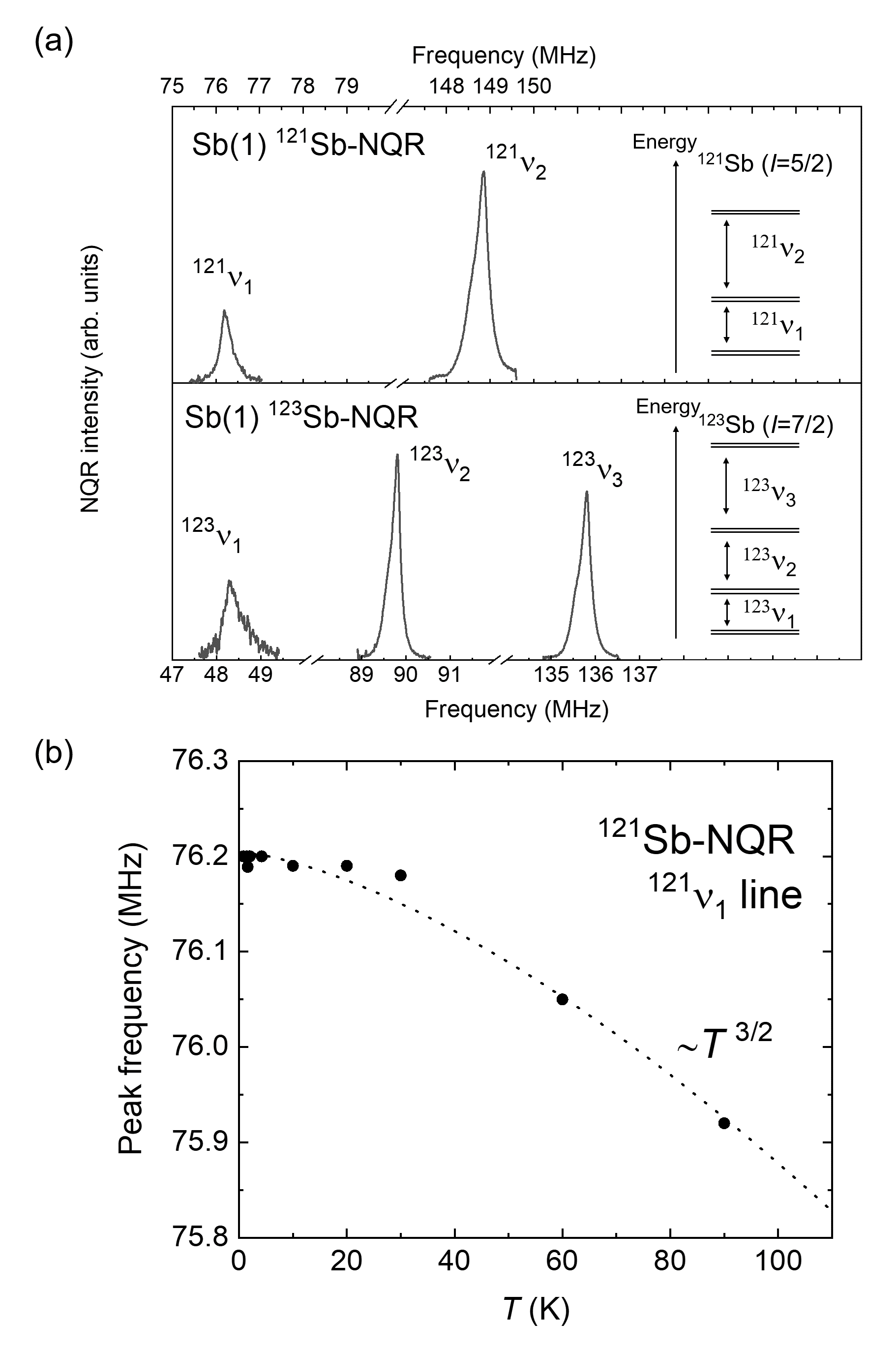}
\caption{(a) NQR spectra of $^{121}$Sb (top) and $^{123}$Sb (bottom) for the Sb(1) site of CaSb$_2$. These are obtained at 4.2 K except for the $^{123}\nu_1$ line, which was measured at 3 K to enhance the intensity. Inset: Schematic of nuclear spin levels split by $\mathcal{H}_Q$. (b) $T$ dependence of the peak frequency of the $^{121}\nu_1$ line. The dotted curve is the result of the fitting with $T^{3/2}$ behavior.}
\label{spectra}
\end{figure}

We show the electronic band structure and DOS in Fig.~2(a). The color in the band structure indicates the extent of the contribution from each Sb site. As we explained earlier, the band calculation with SOC suggested that CaSb$_2$ has Dirac line nodes on the surfaces (at $b^*=\pm\pi/b$) of the Brillouin zone (BZ), which is shown in Fig. 2(b). Here, we took the $c^*$ axis as the $\mathrm{\Gamma Z}$ axis, which is different from the notation used in Ref. 22 (the $b^*$ axis is along the $\mathrm{\Gamma Z}$ axis). Along the YC line, there are two bands below $E_{\mathrm{F}}$ in Fig. 2(a), but along the YA and YE lines we can see only one line. This means that two bands are almost degenerate along these lines in the momentum space due to the line nodes located very close to these lines. Importantly, these nodal lines cross $E_{\mathrm{F}}$. The corresponding Fermi surfaces which mainly originate from the Sb(1) site are shown in red in Fig. 2(c). They are coaxial, two deformed cylinders and touch each other at points in the region indicated by the green ovals in Fig. 2(c). The topological line nodes cross $E_{\mathrm{F}}$ at these points. The precise locations of those line nodes are given in Ref. 22. It should be noted that another Fermi surface not related to the line nodes also exists (blue). 

Figure 3(a) shows the $^{121/123}$Sb-NQR spectra at the Sb(1) site in the normal state. The spectrum of $^{123}$Sb $\nu_1$ line corresponding to the $\pm\frac{1}{2}\leftrightarrow\pm\frac{3}{2}$ transition was measured at 3 K and the other spectra were measured at 4.2 K. The nuclear quadrupolar Hamiltonian is described as
\begin{equation}
\mathcal{H}_Q=\frac{h\nu_Q}{6}\left\{(3I_z^2-I^2)+\frac{1}{2}\eta\ (I_+^2+I_-^2)\right\},
\end{equation}
where $h$ is Plank's constant, $\nu_Q=eQ|V_{zz}|/6I(2I+1)h$ is the NQR coupling constant,  $\eta=|V_{xx}-V_{yy}|/|V_{zz}|$ is the asymmetric parameter, and $V_{ii}$ is the electric field gradient (EFG) along the $i$-axis ($i=x,\ y,\ z)$. The $z$ axis is defined as the principal axis of the EFG tensor with the largest eigenvalue.
 As shown in the inset of Fig. 3(a), the nuclear spin levels split into three and four levels by $\mathcal{H}_Q$ for the $I=5/2$ and $I=7/2$ nuclei, respectively. From the observed peaks, $\nu_Q$ and $\eta$ values are evaluated to be 74.7 MHz and 0.135 for $^{121}$Sb and 45.35 MHz and 0.135 for $^{123}$Sb, respectively. These values are consistent with the band structure calculations performed using WIEN2k, which generate the values of $\nu_Q$ and $\eta$ for $^{121}$Sb(1) as 80.03 MHz and 0.131, and those for $^{123}$Sb(1) as 48.57 MHz and 0.131. Note that the calculated values of $\nu_Q$ and $\eta$ for $^{121}$Sb(2) are 103.31 MHz and 0.210, and those for $^{123}$Sb(2) are 62.70 MHz and 0.210, which indicates that the signals of the Sb(2) site are well separated from those of the Sb(1) site. We tried to observe the Sb(2) NQR spectra by frequency sweeping over several tens MHz centered at the calculated frequencies, but could not observe them. This seems due to the broadening of the NQR spectrum ascribed to the larger value of $\eta$. In fact, the observed NMR peak positions under the magnetic fields are consistent with the WIN2K calculations; the peaks for Sb(2) are much broader than those for Sb(1). Each peak of the Sb(1) NQR spectrum has an asymmetric shape, which can be attributed to the distribution of $\eta$. This is because the shoulder in the $\nu_1$ line is located on the right of the main peak while those in $\nu_2$ and $\nu_3$ are on the left. If $\nu_Q$ had a distribution, each spectrum would have a similar shape with the shoulder always on the same side, and this is not the case. We actually confirmed that the observed NQR spectra were reproduced from $\mathcal{H}_Q$ by assuming a distribution in $\eta$ of approximately 10$\%$. 

Figure 3(b) shows the $T$ dependence of the peak frequency of the $^{121}$Sb $\nu_1$ line. It roughly follows a $T^{\frac{3}{2}}$ behavior, which is a typical behavior in metallic compounds and mainly arises from the $T$ dependence of the lattice constants or lattice vibration.\cite{metalnq}

\begin{figure}
\centering
\includegraphics[clip, width=9.1cm]{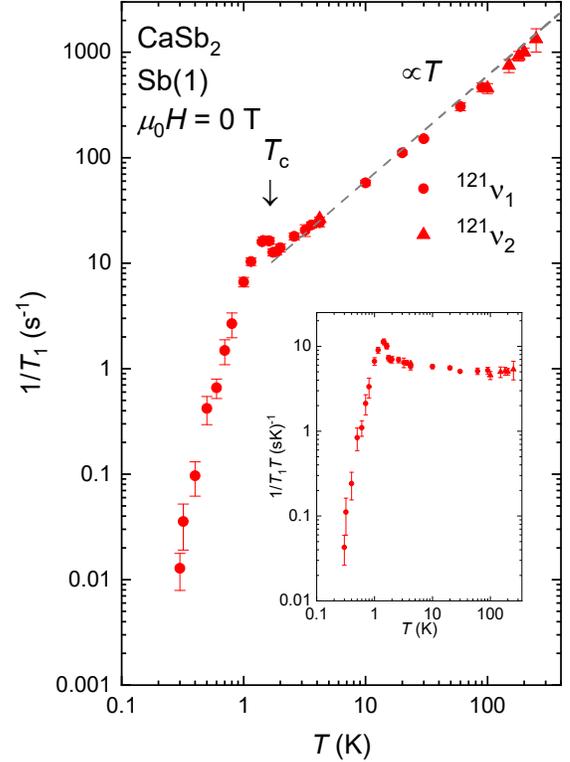}
\caption{$T$ dependence of 1/$^{121}T_1$ measured at the peak frequencis of $^{121}\nu_1$ (filled circles) and $^{121}\nu_2$ (filled triangles) lines in the log-log scale. The dashed line represents the Korringa relationship. Inset: $T$ dependence of $1/^{121}T_1T$.}
\label{t1}
\end{figure}

To investigate low-energy spin excitation in the normal and SC states, we measured $1/T_1$. Figure \ref{t1} shows the $T$ dependence of 1/$^{121}T_1$ measured at the peak frequencies of the $^{121}$Sb $\nu_1$ and $\nu_2$ spectra. We also measured $1/^{123}T_1$ of $^{123}$Sb at 4.2 K to determine the relaxation process. The obtained value of  $^{123}T_1/^{121}T_1$ is 3.15, which is close to the square of the gyromagnetic ratio of the two isotopes $\left(^{121}\gamma/^{123}\gamma\right)^2=3.41$. On the other hand, it is far from $\frac{3(2\cdot^{121}I+3)}{10(2\cdot^{121}I-1)(^{121}I)^2}\left[\frac{3(2\cdot^{123}I+3)}{10(2\cdot^{123}I-1)(^{123}I)^2}\right]^{-1}\left(^{121}Q/^{123}Q\right)^2= 1.50$,\cite{obata} which is the value expected when the electronic quadrupole relaxation process is dominant. This result indicates that $1/T_1$ is governed by the magnetic interaction through  hyperfine coupling. 
Above $T_{\mathrm{c}}$, $1/T_1$ roughly follows the Korringa law (proportional to $T$), which is a typical behavior of a conventional normal metal. Note that $1/T_1T$ gradually increases below 100 K as shown in the inset of Fig. 4. This would be due to the weak correlation among the electrons. Since it has been reported that the nodal-line properties appear in the form of power law dependence on $T$ $(1/T_1\propto T^n \ \mathrm{with}\  n>1)$ with a logarithmic factor of $T$,\cite{maebashi} the Korringa-like behavior indicates that the nodal-line properties expected from the degeneracy of two bands are masked by the contribution from Fermi-surface portions away from nodes. In fact, the value of $1/^{121}T_1T$ is approximately 6 (sK)$^{-1}$, which is several orders of magnitude larger than those of the systems in which nodal properties are observed by NMR/NQR measurements.\cite{tap, tino} The large value of $1/T_1T$ reflects the contribution from ordinary electrons, which is consistent with the band structure calculation presented in Fig. 2.

In the SC state, 1/$T_1$ shows a clear coherence peak just below $T_{\mathrm{c}}=1.7$ K. The maximum value of $1/T_1$ is 1.6 times as large as $1/T_1$ just above $T_{\mathrm{c}}$. Such a large coherence peak cannot be explained unless conventional $s$-wave pairing with a full gap is realized.\cite{peak} Correspondingly, $1/T_1$ decreases exponentially at lower temperatures. This exponential behavior can be directly shown in the Arrhenius plot of $T_1$ against $T_{\mathrm{c}}/T$ in Fig. 5.

To evaluate the magnitude of the SC gap $\Delta(0)/k_{\mathrm{B}}T_{\mathrm{c}}$ from the $T$ dependence of $1/T_1$, a numerical calculation based on the BCS theory was performed. 1/$T_1$ in the SC state (1/$T_{1\mathrm{s}}$) normalized by that in the normal state (1/$T_{1\mathrm{n}}$) is expressed as
\begin{equation}
\frac{T_{1\mathrm{n}}}{T_{1\mathrm{s}}}=\frac{2}{k_{\mathrm{B}}T}\int_0^{\infty} dE\ N_{\mathrm{s}}^2(E)\left[1+\frac{|\Delta(T)|^2}{E^2}\right]f(E)\left[1-f(E)\right],
\end{equation}
where $N_{\mathrm{s}}(E)$ is the quasi particle DOS in the SC state, $\Delta(T)$ is the $T$ dependent energy gap, and $f(E)$ is the Fermi distribution function. The factor $\left[1+\frac{|\Delta(T)|^2}{E^2}\right]$ is related to the coherence effect in the SC state. Following a previous study,\cite{hebel} we considered the energy broadening in $N_{\mathrm{s}}(E)$ by taking the convolution of $N_{\mathrm{s}}(E)$ with a rectangular broadening function whose width and height are $2\delta$ and $1/2\delta$, respectively.\cite{hebel} Using  $\Delta(0)/k_{\mathrm{B}}T_{\mathrm{c}}=1.52$ and $\delta/\Delta(0)=0.35$, the experimental data were well reproduced as shown in the main panel and the inset of Fig. 5. The $\Delta(0)/k_{\mathrm{B}}T_{\mathrm{c}}$ value used in the calculation is close to the value expected from the weak-coupling $s$-wave BCS theory (1.76). This agreement, as well as the clear coherence peak and the exponential decrease, strongly indicate that conventional full-gap superconductivity is realized in CaSb$_2$.

We here compare our results with other experiments. As previously reported,\cite{ikeda} the specific heat shows a broad peak, which indicates the distribution of $T_{\mathrm{c}}$ within the sample. On the other hand, our coherence peak is sharp and the SC behavior of $1/T_1$ can be fitted without assuming the distribution of $T_{\mathrm{c}}$. This is because NQR is a microscopic measurement which is able to extract the electronic state in a certain environment with high selectivity. The difference in the NQR frequency originates from the difference in the local environment surrounding the nucleus. Therefore, $1/T_1$ measured at a fixed frequency is determined by the electronic state with a single $T_{\mathrm{c}}$. 

The first-principles calculation result indicates that CaSb$_2$ has a pair of cylindrical Fermi surfaces along the zone boundary with a symmetry-protected nodal line, and they offer the possibility of topological superconductivity.\cite{ikeda} However, based on our experimental results, it is concluded that the normal state behaves as a conventional metal and the superconductivity is topologically trivial.
Other physical probes sensitive to the Dirac nodes, such as transport and angle-resolved photoemission spectroscopy measurements, may reveal unconventional normal-state behavior. Furthermore, topological non-trivial character of nodal lines may be enhanced by pressure or chemical substitutions, providing a possibility of topological superconductivity.

\begin{figure}
\centering
\includegraphics[clip, width=9.8cm, height=8.4cm]{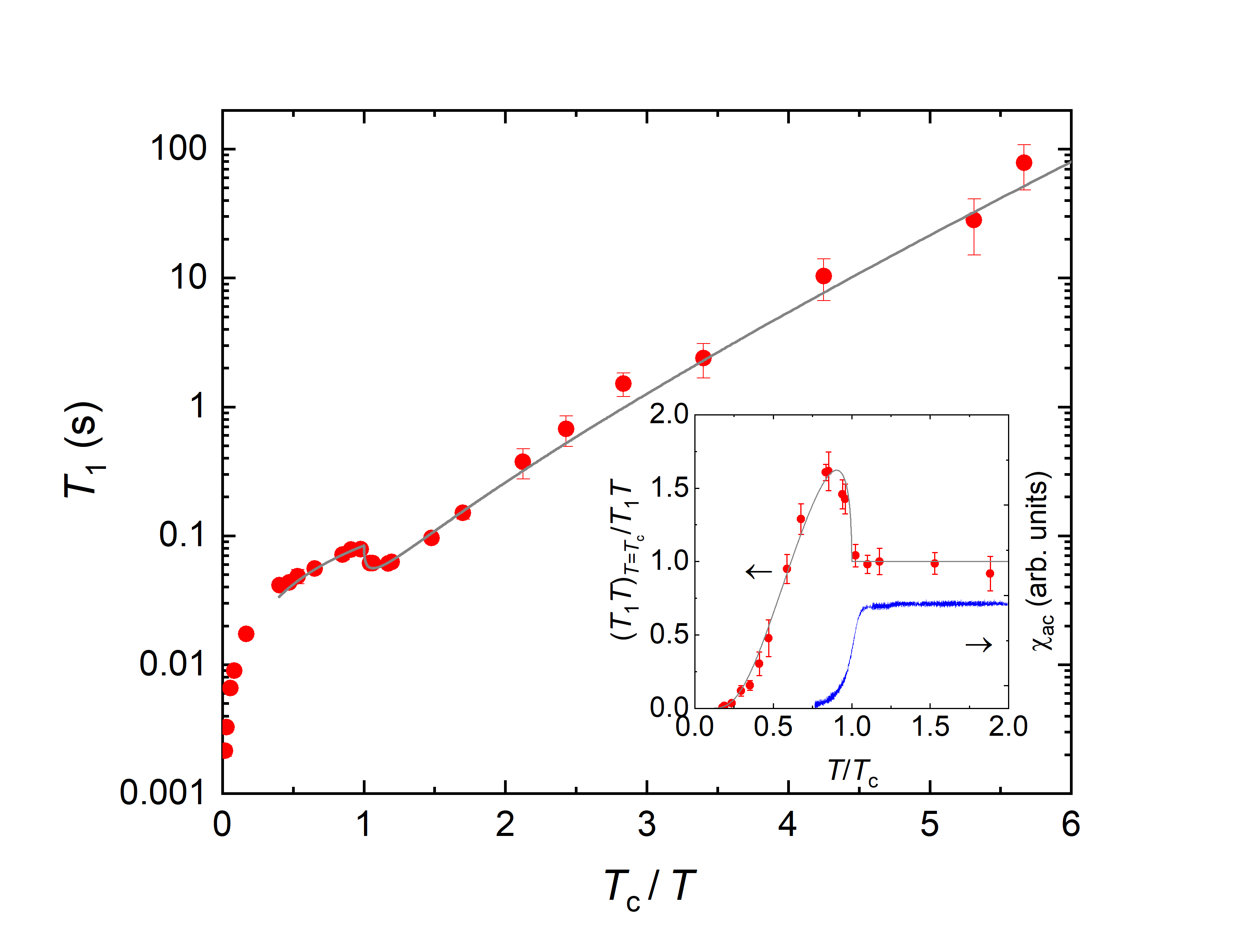}
\caption{Arrhenius plot of $^{121}T_1$ against $T_{\mathrm{c}}/T$ ($T_{\mathrm{c}}=1.7$ K). The gray solid curve is a calculation using the BCS theory [eq. (2)]. Inset: $1/T_1T$ normalized by the value at $T_{\mathrm{c}}$ (red points) and AC susceptibility (blue solid curve) plotted against $T/T_{\mathrm{c}}$. The calculation using the BCS theory is shown in the gray solid curve. }
\label{coherence}
\end{figure}

In conclusion, we performed $^{121/123}$Sb-NQR measurements on the non-symmorphic line-nodal material CaSb$_2$ using a powder sample. The temperature dependence of $1/T_1$ approximately exhibits Korringa behavior above $T_{\mathrm{c}}$, and shows a clear coherence peak just below $T_{\mathrm{c}}$ and an exponential decrease sufficiently below $T_{\mathrm{c}}$.  The first feature indicates that conventional metallic behavior is dominant in the normal state above $T_{\mathrm{c}}$, and the latter two provide strong evidence for an $s$-wave superconductivity. The gap size was evaluated to be $\Delta(0)/k_{\mathrm{B}}T_{\mathrm{c}}=1.52$. By clarifying its SC symmetry, this work will substantially contributes to the understanding of the properties of CaSb$_2$ and those of line-nodal materials.

\vspace{11pt}

\begin{acknowledgment}

We thank T. Hashimoto, A. Yamakage, M. Sato, and T. Oguchi for valuable discussions.
We acknowledge the Research Center for Low Temperature and Materials Sciences, Kyoto University for the stable supply of liquid helium. This work was partially supported by Grant-in-Aid for Scientific Research on Innovative Areas from the Ministry of Education, Culture, Sports, Science, and Technology (MEXT) of Japan, and JSPS Core-to-core program (JPJSCCA20170002). We are also supported by: JSPS KAKENHI Nos. JP15H05852, JP15K21717, JP17H06136, JP20H00130, JP15K21732, JP15H05745, JP20KK0061, JP19H04696, JP19K14657, and JP20H05158. H. Takahashi was supported by Iwadare Scholarship Foundation.

\end{acknowledgment}

\end{document}